# Coherent linking between confocal amplitude image and confocal phase image in dual-comb microscopy

Takahiko Mizuno[1], Takuya Tsuda[2], Eiji Hase[1], Hirotsugu Yamamoto[3], Takeo Minamikawa[1], and Takeshi Yasui[1]

[1]Institute of Post-LED Photonics (pLED), Tokushima University, 2-1 Minami-Josanjima, Tokushima, Tokushima 770-8506, Japan

[2]Graduate School of Advanced Technology and Science, Tokushima University, 2-1 Minami-Josanjima, Tokushima, Tokushima 770-8506, Japan

[3]Center for Optical Research and Education, Utsunomiya University, 7-1-2, Yoto, Utsunomiya, Tochigi, 321-8585, Japan

## Abstract

This paper presents a coherent linking approach between confocal amplitude and confocal phase images acquired using dual-comb microscopy (DCM). DCM combines the advantages of confocal laser microscopy and quantitative phase microscopy, offering high axial resolution and scan-less imaging capability. By exploiting the coherence between confocal amplitude and phase images within the same DCM system, we accurately determine the number of phase unwrapping iterations, eliminating phase wrapping ambiguity. The method is demonstrated with samples having micrometer-range optical thickness and nanometer-scale surface roughness. The results showcase an expanded axial dynamic range, ranging from micrometers to millimeters, while maintaining nanometer-level axial resolution. This coherently linked DCM imaging technique enables the simultaneous acquisition of absolute phase information, enhancing its potential for high-axial-resolution imaging in a wide range of applications.

## 1. Introduction

Two representative optical microscopes that use laser light are confocal laser microscopy (CLM) and quantitative phase microscopy (QPM). CLM [1-3] is a powerful tool for optical-sectioning two-dimensional (2D) imaging or three-dimensional (3D) imaging with depth selectivity. Since CLM enables us to extract a small volume fraction of a sample in the vicinity of a focal point based on confocal optics, 2D or 3D scanning of the focal spot gives 2D or 3D images of thick transparent objects or micrometer-unevenness reflective objects; however, the axial resolution remains around micrometer order due to the confocality. Conversely, QPM [4-6] has been used to visualize thin transparent objects or nanometer-unevenness reflective objects with the phase resolution of milliradians order corresponding to the axial resolution of nanometer order. In QPM, although phase shifts in light passing through or reflecting the object are converted to brightness changes in the image by interference, the phase value repeats every $2\pi$ rad (namely, phase wrapping), making it difficult to visualize thick specimens or micrometer-unevenness reflective objects due to the phase wrapping ambiguity. Although CLM and QPM are complementary to each other from the viewpoint of axial resolution, it is difficult to integrate them with a single experimental setup because of different principles of operation.

Recently, dual-comb microscopy (DCM) [7-13] has attracted attentions as an imaging modality to integrate CLM and QPM together with the scan-less imaging capability. In DCM, by using an optical frequency comb (OFC) [14-16] as an optical carrier of amplitude and phase with a large number of discrete frequency channels, image pixels to be measured are spectrally encoded into OFC modes via space-to-wavelength transformation (namely, spectral encoding or SE); then, they are decoded all at once from the mode-resolved spectrum of the image-encoded OFC acquired by dual-comb spectroscopy (DCS) [17-20], based on one-to-one correspondence between images pixels and OFC modes. The resulting confocal amplitude and phase images have been effectively applied for the surface topography of a nanometer-scale step-structured sample [7], the non-staining visualization of culture fixed cells [7], and the real-time movie of moving objects [7,12,13].

In previous studies of DCM, the confocal amplitude and the confocal phase image were separately acquired. However, since these two images are acquired by the same light source and the same experimental setup, they can be coherently linked to each other. If confocal amplitude and phase images are connected, it enables optical imaging featuring wide dynamic range of axial direction based on the complementarity between the confocal amplitude image and the confocal phase image. In this article, we demonstrate coherent linking between confocal amplitude image and confocal phase image in DCM.

## 2. Principle of operation

We first consider to measure a sample with internal structure (for example, cell cultured on a glass substrate) by CLM and QPM, independently. An upper part of Fig. 1(a) shows a behavior of axial profile measured by CLM. Two peaks appear at axial positions corresponding to boundaries of internal structures, and their separation gives an optical thickness of internal structure in the sample. When each boundary is sufficiently thin, width of each peak profile is corresponding to the confocal profile of CLM, which limits the axial resolution (typically, ~ several μm depending on the confocal optics). A lower part of Fig. 1(a) shows a behavior of axial profile measured by QPM, in which the red and blue plots indicate the behavior of the phase signal returned from the first and the second boundaries, respectively. These signals repeat phase wrapping with respect to the axial position and gives relative phase values corresponding to boundaries of internal structures. Those relative phase values can achieve an axial resolution of sub-wavelength or nanometer. The absolute position of each boundary or optical thickness of internal structure can be obtained by measuring the number of phase wrapping. Unfortunately, the number of their phase wrapping iterations is too large to determine it correctly, leading to the phase wrapping ambiguity. More importantly, two phase signals from different boundaries [see red and blue plots in the lower part of Fig. 1(a)] are interfered with each other, disturbing the phase information. In this way, it is difficult to apply QOM for the sample with internal structures.

We next consider to measure the same sample by DCM. An upper part of Fig. 1(b) shows a behavior of axial profile in the confocal amplitude image measured by DCM. The axial profile is similar to that in CLM as shown in the upper part of Fig. 1(a). A lower part of Fig. 1(b) shows a behavior of axial profile in the confocal phase image measured by DCM. Compared with the axial profile of QPM [see the lower part of Fig. 1(a)], the confocality of DCM limits the range of phase wrapping appearing within the confocal amplitude profile [see the upper part of Fig. 1(b)]. This suppresses the overlapping of two phase signals from different boundaries (see red and blue plots). More importantly, since the confocal amplitude is coherently linked with the confocal phase images due to use of the same light source and the same experimental setup, the peak of confocal amplitude profile can be used as an indicator to determine the number of phase wrapping iterations. In other words, the coherently linking between the confocal amplitude and phase images can give the absolute phase vales calculated from the relative phase value and the number of phase wrapping iterations, and hence eliminate the phase wrapping ambiguity. In this way, the coherently linking largely expands the axial range over micrometer to millimeter order while maintaining the nanometer-order axial resolution.

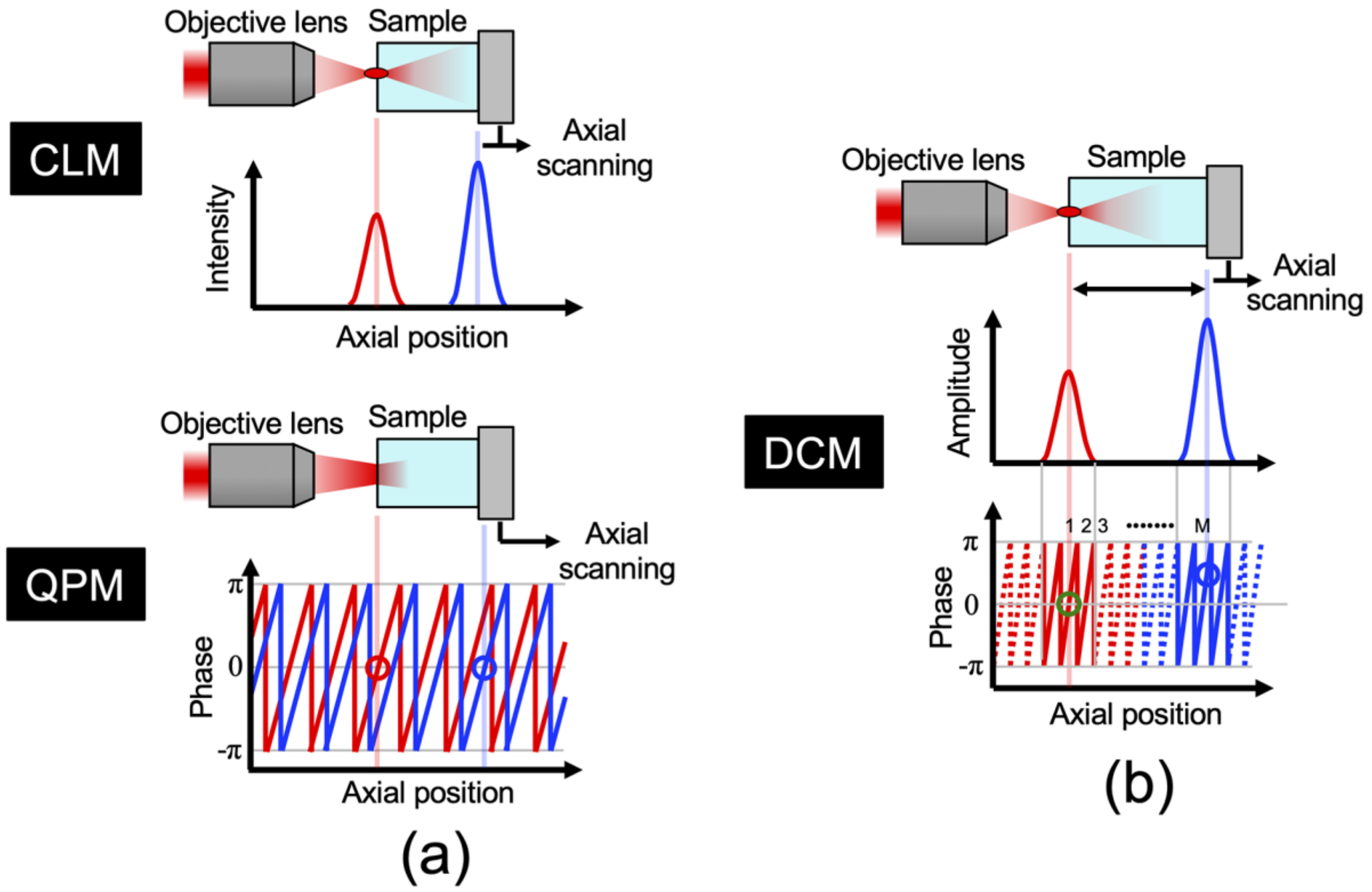

Fig. 1. Principle of operation. (a) Axial profile of a sample with internal-structure measured by CLM (upper part) and QPM (lower part). (a) Axial profile of the same sample in confocal amplitude image (upper part) and confocal image (lower part) measured by DCM.

## 3. Experimental setup

Figure 2 depicts the experimental setup of the DCM. As the comprehensive details of the experimental setup can be found elsewhere [12], we will now provide a concise overview of the key components. An optical beam emanating from a signal OFC (center wavelength = 1560 nm, spectral range = 1545~1575 nm, mean output power = 125 mW, and repetition rate $f_{rep1}$ = 100,388,730 Hz), referred to as the "signal OFC beam," passed through an optical bandpass filter (BPF, passband wavelength = 1554-1566 nm) and a beam expander composed two lenses (L1 and L2); then, it was split into two arms: a reference arm to generate a reference OFC beam and a two-dimensional spectral encoding (2D-SE) arm to produce an image-encoded OFC beam. This separation was achieved using a 50:50 beam splitter (BS). The image-encoded OFC beam passed through the BS and was directed into a 2D-SE optical system comprising a virtually imaged phased array (VIPA) with a free spectral range of 15.1 GHz and a finesse of 110, along with a diffraction grating (groove density = 1200 grooves/mm, efficiency = 90%). The beam then underwent 2D spectral dispersion of the signal OFC modes, facilitated by a pair of lenses (L3 and L4, focal length = 150 mm) and a dry-type objective lens (OL, numerical aperture = 0.95). This results in the generation of a 2D focal spot array consisting of 14,800 spots on the sample surface. The sample was placed at a translation stage for the axial scanning. When the image-encoded OFC beam interacted with the sample, the image contrast was encoded onto the amplitude and phase spectra of the 2D spectrograph through reflection, absorption, scattering, and/or phase changes. Subsequently, the 2D spectrograph of the image-encoded OFC passed through the same optical system in the opposite direction, causing each wavelength component of the spectrograph to spatially overlap once again, effectively reconstituting the image-encoded OFC beam. Following this, the image-encoded OFC beam was combined with the reference OFC beam, ensuring a time separation of 6.2 ns, and then coupled into a single-mode fiber coupler (SMF-C) for DCS. The SMF-C functions as a confocal pinhole of DCM.

An optical beam emanating from a local OFC (center wavelength = 1560 nm, spectral range = 1545~1575 nm, mean output power = 15 mW, $f_{rep2}$ = 100,389,709 Hz, $\Delta f_{rep} = f_{rep2} - f_{rep1}$ = 979 Hz), referred to as the "local OFC beam," passed through polarization controller (PC) and another optical bandpass filter (BPF2, passband wavelength = 1554-1566 nm); then, it was spatially overlapped with the combined image-encoded and reference OFC beam by the SMF-C. For optimal visibility of the interferogram in the time domain, we maintained an optical power ratio of 1:1 between the local OFC beam and the combined image-encoded and reference OFC beam. After optical amplification by a custom-made erbium-doped fiber amplifier (EDFA), the interferogram signal was detected by a fast photodetector (PD) with a bandwidth ranging from DC to 1.2 GHz. This detector was connected to a low-noise amplifier (AMP) with a bandwidth spanning from 1 kHz to 100 MHz. The amplified electric signal was then acquired by a digitizer for further analysis and processing.

The acquired interferogram was separated into the image-encoded OFC interferogram and the reference OFC interferogram with a time window of a half of $1/f_{rep1}$. A Fourier transform of these two interferograms provides their corresponding amplitude and phase spectra. Optical passband of BPF1 and BPF2 was set to avoid aliasing effect in DCS. We calculated an amplitude-ratio spectrum and a phase-difference spectra between the image-encoded OFC and the reference OFC. Each data plot of the amplitude and phase spectra was spatially mapped to the confocal amplitude and phase images, ensuring a one-to-one correspondence between 2D image pixels and OFC modes.

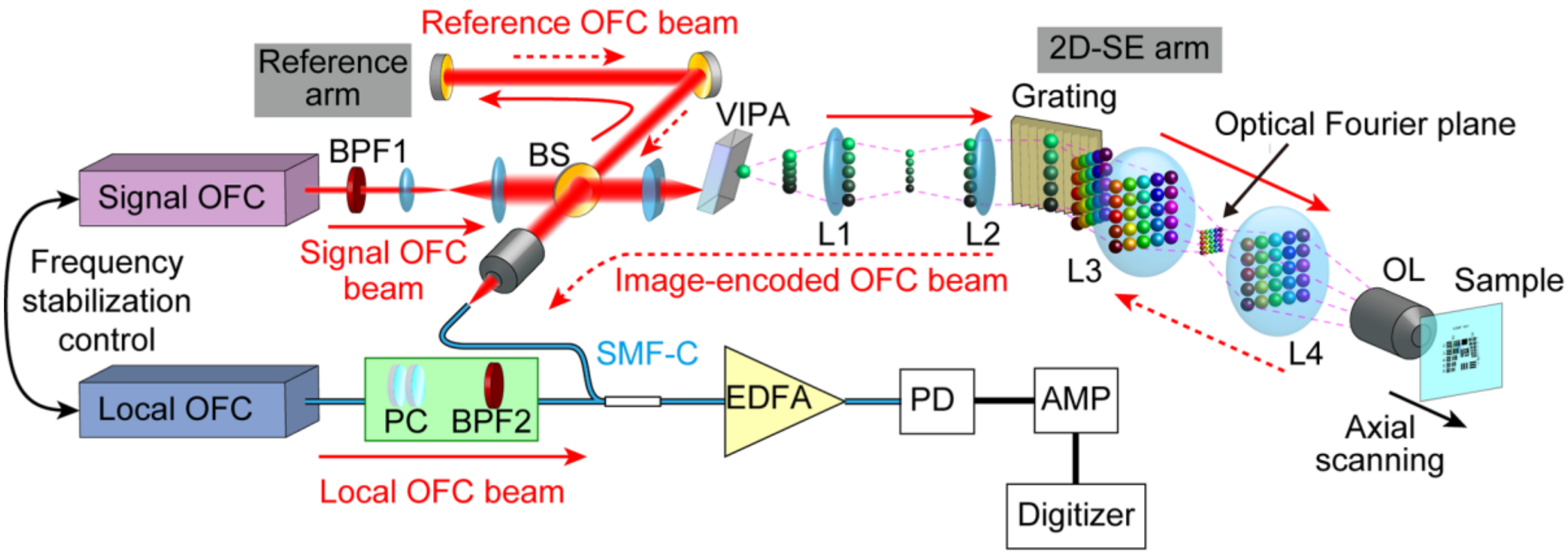

Fig. 2. Experimental setup. BPFs, optical bandpass filters; BS, beam splitter; VIPA, virtually imaged phased array; L1, L2, L3, L4, lenses; OL, objective lens; PC, polarization controller; EDFA, erbium-doped fiber amplifier.

## 4. Results

To determine the number of phase unwrapping in confocal phase imaging from the peak axial position of confocal amplitude imaging, both confocal amplitude imaging and confocal phase imaging must exhibit high reproducibility of axial profile, respectively. We first evaluated the reproducibility in both imaging. A gold mirror was used as a reflective sample without internal structures. We extracted a value of a center pixel in the image, which has one-to-one correspondence with the mode of the image-encoded OFC at a wavelength $\lambda$ of 1560 nm. We repeated confocal amplitude imaging and confocal phase imaging of this sample with respect to the axial position (axial step = 62.5 nm) four times. Figure 3(a) shows a comparison of axial profile among confocal amplitude images. We confirmed that the confocal axial profiles exhibit Gaussian shapes, and their full width at half maximum (FWHM), representing the confocal axial resolution, is 8.2±0.11 μm (mean ± standard deviation). The peak position of the axial position was determined by curve fitting analysis with a Gaussian function, and the determined peak positions have a standard deviation of 51 nm. Compared with the phase wrapping period (= $\lambda/2$ = 780 nm), the standard deviation was sufficiently small.

Conversely, Fig. 3(b) shows a comparison of axial profile among confocal phase images. We observed that phase wrapping was repeated within the axial range of the confocal axial profile and each profile moderately overlaps with one another. At the axial position of 0 μm, the variation of relative phase values was found to be 29 nm, as indicated by the standard deviation. To confirm whether the confocality of DCM limits the range of phase wrapping appearing within the confocal amplitude profile, we also acquired the axial profiles of the confocal phase image outside the range of the confocal amplitude profile. Figure 3(c) shows the axial profile of confocal phase image around the axial position of -150 μm [not shown in Fig. 3(a)]. In contrast to the behavior of Fig. 3(b), the regular behavior of phase wrapping has disappeared, and instead, random phase noise is observed. In this way, both confocal amplitude imaging and confocal phase imaging have the reproducibility of axial profile sufficient to determine the number of phase unwrapping.

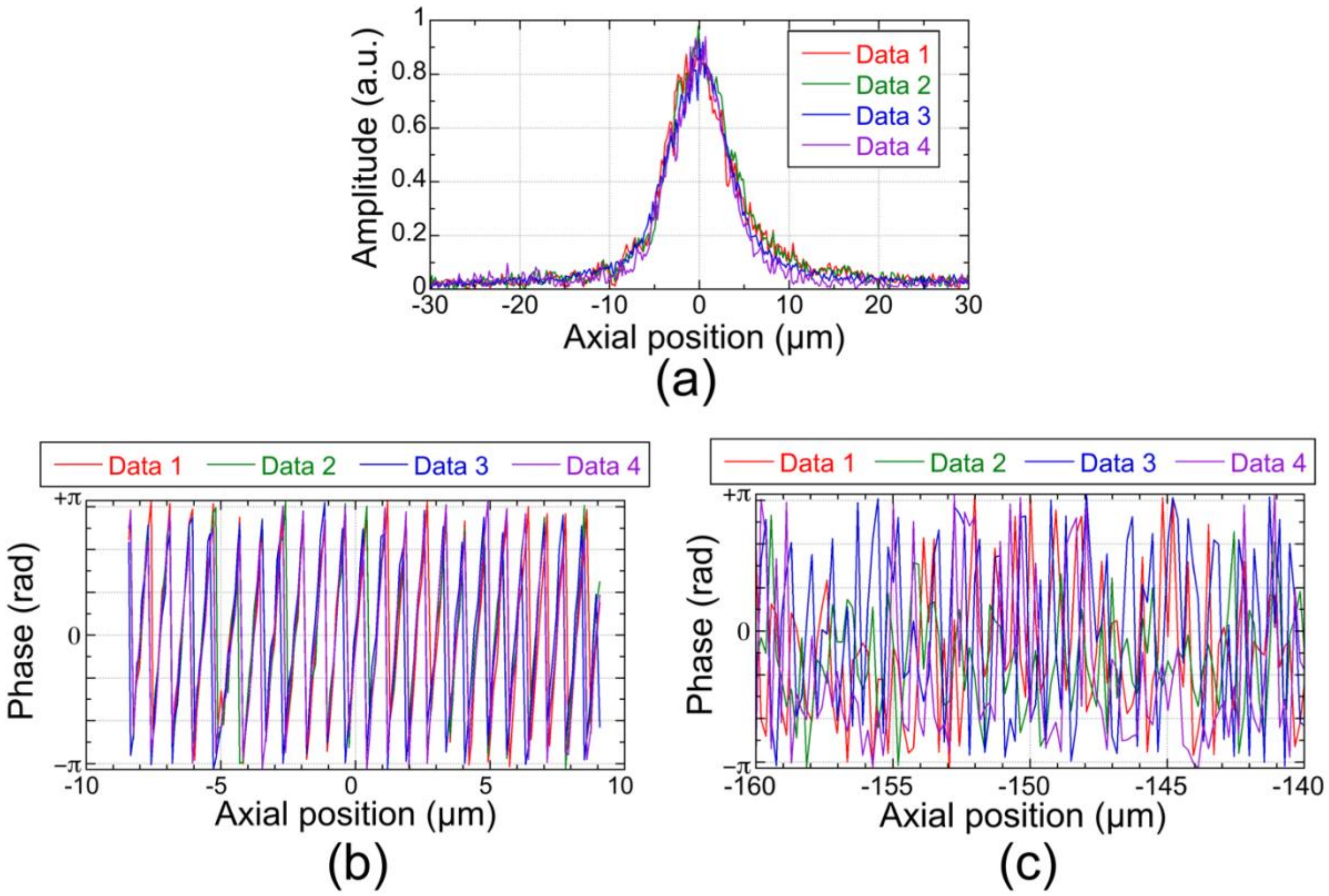

Fig. 3. Basic performance of JM-DCSP. Reproducibility of axial profile in (a) confocal amplitude imaging and (b) confocal phase imaging. Axial profile in confocal amplitude imaging outside the confocality.

We next performed the confocal amplitude imaging and confocal phase imaging of a sample with internal structure with respect to the axial position. Figure 4(a) shows a schematic drawing of the sample. We put a cover glass (thickness = 130~170 μm) on a 1951 USAF resolution test chart with a negative pattern (Edmond Optics, Barrington, NJ, USA, #38–256, spatial frequency: 1.00 lp/mm ~ 228 lp/mm). The presence or absence of a reflective coating on the test chart imparts image contrast, reflecting the differences in reflectivity and surface roughness. Figure 4(b) shows an axial profile in the confocal amplitude imaging of this sample. We here defined the axial position of the first peak to be 0 μm. The first peak at $z_2$ (= 0.0 μm) reflects a surface of the cover glass whereas the second peak at $z_4$ (= +94.0 μm) indicates a boundary between the cover glass and the test chart. No signal appeared at $z_1$ (= -31.4 μm), $z_3$ (= +38.4 μm), and $z_5$ (= +138.6 μm). Figure 4(c) shows confocal amplitude images at $z_1$, $z_2$, $z_3$, $z_4$, and $z_5$ (image size = 85 μm by 90 μm). The plane image resulting from the surface reflection of the cover glass appeared at $z_2$ whereas the test-pattern image resulting from the reflection of the test-chart coating surface appeared at $z_4$. No images appeared at $z_1$, $z_3$, and $z_5$. The optical thickness of the cover glass was determined to be 94 μm from the separation between two peaks in the axial profile in Fig. 4(b). However, the uncertainty is limited by the confocal axial resolution of 8.2 μm in DCM.

Figure 4(d) presents the confocal phase images at $z_1$, $z_2$, $z_3$, $z_4$, and $z_5$ (image size = 85 μm by 90 μm). Each image demonstrates the spatial distribution of relative phase values, similar to previous DCM studies. However, a significant difference from prior research lies in the simultaneous determination of the number of phase unwrapping iterations based on the comparison between the multiple phase wrapping of the confocal amplitude image and the peak position of the confocal amplitude profile. The phase images at $z_1$, $z_3$, and $z_5$ show random phase noise due to no signal. The phase image at $z_2$ exhibits a horizontal gradient in phase, indicating an inclination of the cover glass surface within a single wavelength along the horizontal direction. The relative phase ($\phi_{z_2}$) of the center pixel at $z_2$ was set to be 0.0 rad. Also, the number of phase wrapping iterations at $z_2$ ($M_{z_2}$) was set to be 0. The phase image at $z_4$ reflects the surface unevenness of test pattern caused by the presence or absence of the reflective coating. The phase difference across the test pattern coating was 0.53 rad corresponding to the surface unevenness of 66 nm, which is in good agreement with the surface unevenness of 60 nm measured by the atomic force microscopy (Hitachi High-Tech, AFM5500M, axial repeatability < 1 nm). The relative phase ($\phi_{Z_4}$) of the center pixel at $z_4$ was 1.27 rad whereas the number of phase wrapping iterations at $z_4$ ($M_{Z_4}$) was determined to be 121. The optical thickness ($n_g d$) of the cover glass is given by

$$n_g d = \lambda\left[\left(M_{Z_4} - M_{Z_2}\right) + \frac{\phi_{Z_4} - \phi_{Z_2}}{2\pi}\right] = 1.56\left[121 + \frac{1.27}{2\pi}\right] = 189.08\ \mu\text{m}, \quad (1)$$

where $n_g$ and $d$ are a group refractive index (= 1.53) and a geometrical thickness (= 130 ~ 170 μm in specification) of the cover glass, respectively. This optical thickness is in moderate agreement with the specification of the cover glass.

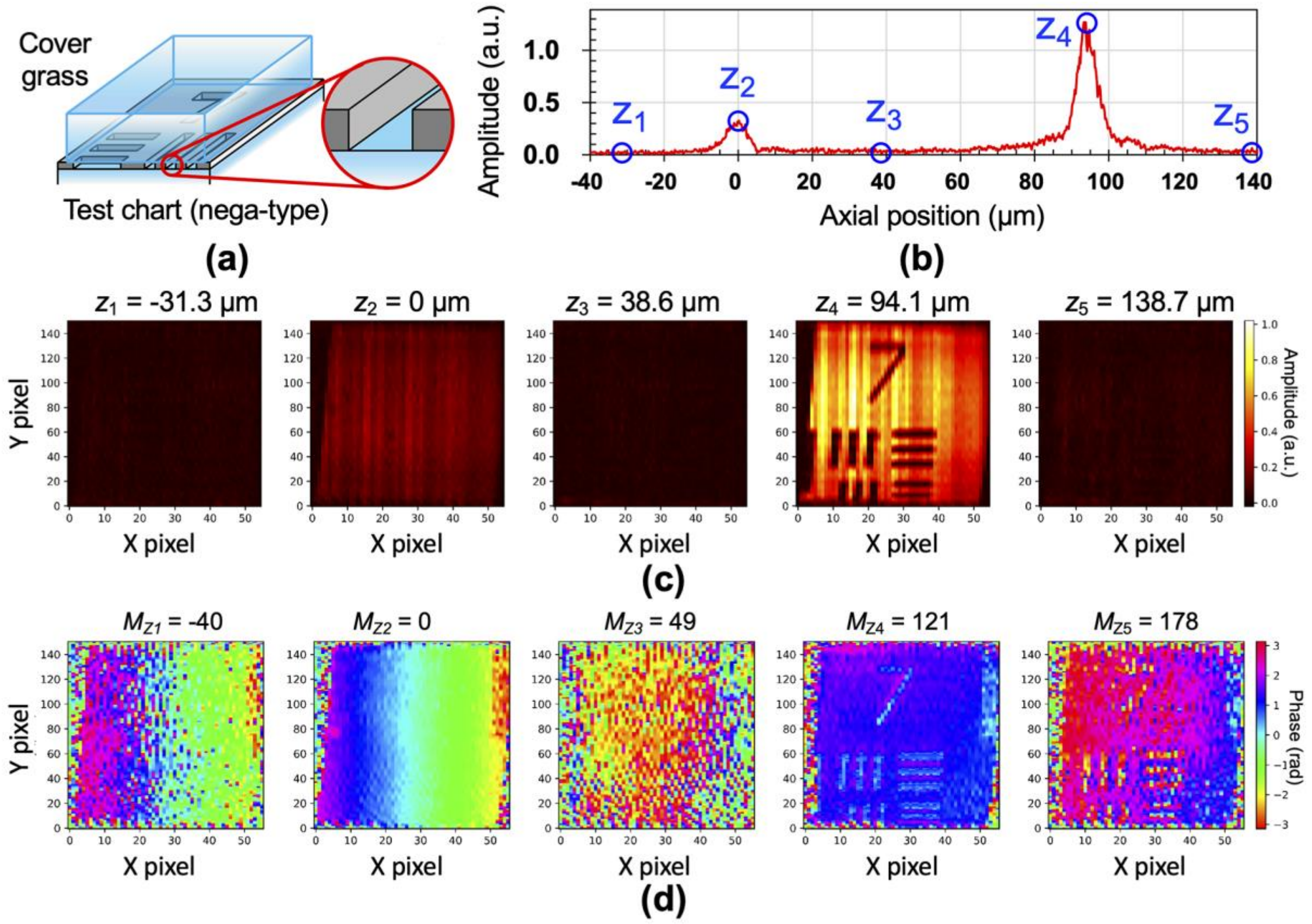

Fig. 4. Coherently-linked confocal amplitude and phase imaging. (a) Schematic drawing of a sample. (b) Axial property of confocal amplitude imaging. (c) A series of confocal amplitude images at $z_1$, $z_2$, $z_3$, $z_4$, and $z_5$. (d) A series of confocal phase

images at $z_1$, $z_2$, $z_3$, $z_4$, and $z_5$.

## 5. Discussion

As shown in Fig. 3, the confocality of DCM limits the range of phase wrapping iterations appearing within the confocal amplitude profile. However, multiple phase wrapping iterations was still observed [see Fig. 3(b)] because the confocal axial resolution of confocal amplitude imaging is larger than the phase wrapping period (= $\lambda/2$) of confocal phase imaging. If the confocal axial resolution is equal to the phase wrapping period, only a single phase slope appears in the axial profile of confocal phase imaging, as shown in Fig. 5. This is because the confocality of DCM suppresses to the multiple phase wrapping iterations and extracts the single phase slope. The resulting phase imaging with the confocal axial selectivity would be free of the phase wrapping ambiguity. In in the demonstration above, two boundaries of internal structure in the sample should be separate by the confocal axial resolution of DCM (= 8.2 μm) at least; however, the consistency between the confocal axial resolution and phase wrapping period enables the separation of phase signals returning from multiple boundaries with sub-μm order. The consistency can achieve the seamless connection between the confocal amplitude and phase images in more simple way.

We here discuss a space to further enhance the confocal axial resolution by improvement of the optical system. The theoretical value of confocal axial resolution ($\Delta z$) is given by

$$\Delta z = \frac{0.68\lambda}{n - \sqrt{n^2 - NA^2}} , \quad (2)$$

where $n$ and $NA$ are a refractive index and a numerical aperture of the objective lens, respectively. For example, if an oil-immersion, high-NA objective lens (NA = 1.4, n =

1.4) is used in a sufficiently small confocal pinhole, Δz is achieved to 0.75 μm. In other words, the seamless connection between the confocal amplitude and phase images will be achieved based on the consistency between the confocal axial resolution and phase wrapping period.

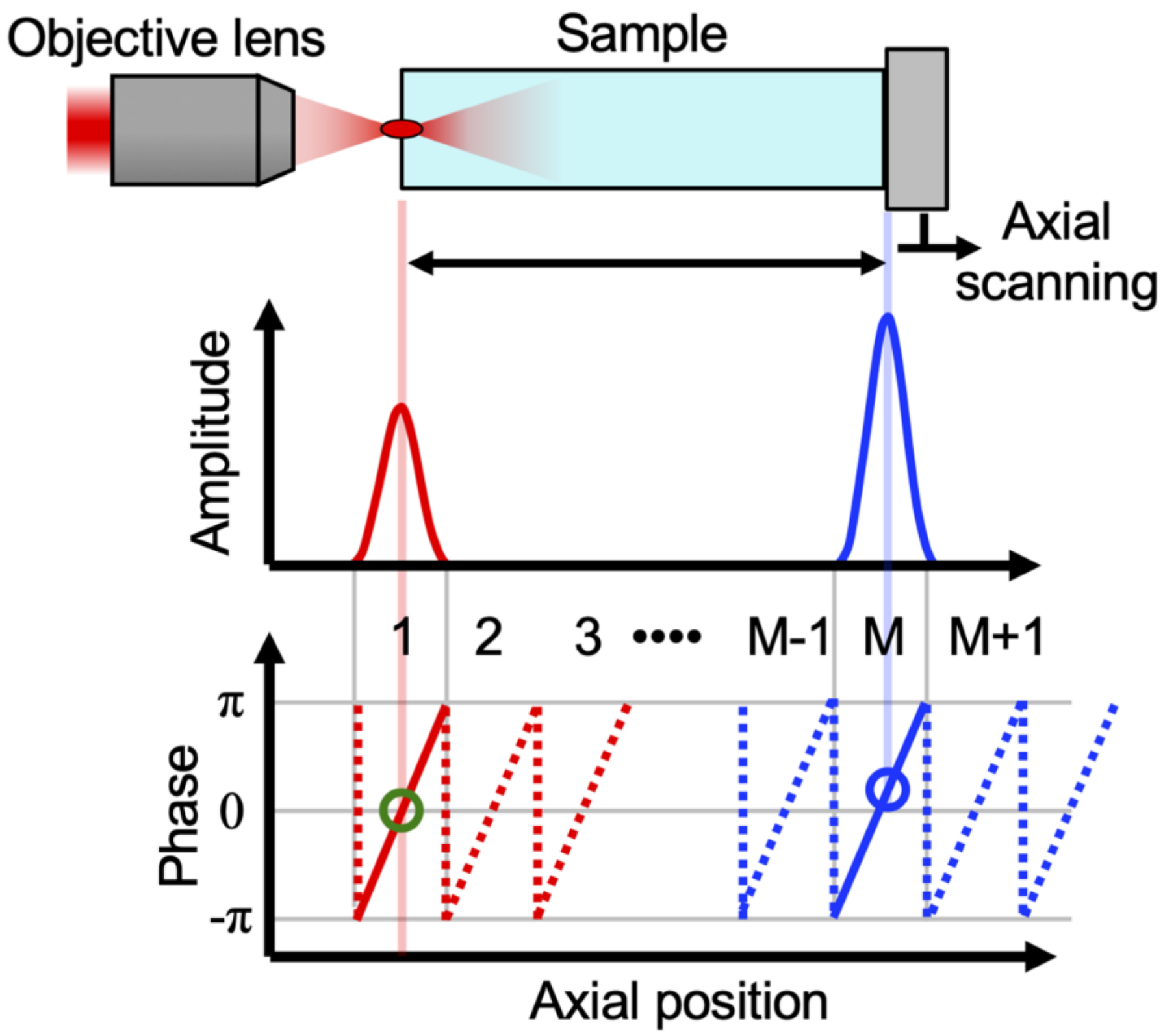

Fig. 5. Extraction of a single phase slope in confocal phase imaging by matching confocal axial resolution with phase wrapping period.

## 6. Summary

In this article, we proposed an approach to coherently link both confocal amplitude and confocal phase images, which can be simultaneously acquired using the same DCM system. The obtained images exhibited high reproducibility along the axial direction. By accurately determining the number of phase unwrapping iterations in the confocal phase image based on the peak of the confocal amplitude profile, we were able to obtain absolute phase images in addition to providing relative phase information through the confocal phase image. We confirmed the effectiveness of our proposed method through measurements of samples with optical thicknesses in the micrometer range and surface roughness in the nanometer range. As a result, our approach enables DCM to achieve a broad axial range spanning from micrometers to millimeters with nanometer-level axial resolution, significantly expanding the axial dynamic range.

Our proposed method offers a valuable means to obtain absolute phase information from confocal amplitude and confocal phase images simultaneously acquired using DCM. This advancement enhances the capabilities of DCM in providing high-resolution imaging over a wide axial range, from micrometers to millimeters, with nanometer-level precision, thereby expanding its potential applications.

## Acknowledgments

This work was supported by the Exploratory Research for Advanced Technology (JPMJER1304), Japan Society for the Promotion of Science (18H01901, 18K13768, 19H00871, 22H00303), Cabinet Office, Government of Japan (Subsidy for Reg. Univ. and Reg. Ind. Creation), Nakatani Foundation for Advancement of Measuring Technologies in Biomedical Engineering, Research Clusters program of Tokushima University (2201001).

## Author declarations

### Conflict of Interest

The authors have no conflicts to disclose.

**Author Contributions**

**Takahiko Mizuno**: Data curation (equal); Formal analysis (lead); Investigation (lead); Methodology (equal); Software (lead); Writing – original draft (equal); Writing – review & editing (equal). **Takuya Tsuda**: Data curation (equal); Formal analysis (supporting); Investigation (supporting). **Eiji Hase**: Formal analysis (supporting); Methodology (equal). **Hirotsugu Yamamoto**: Conceptualization (equal); Funding acquisition (supporting); Methodology (equal); Supervision (supporting). **Takeo Minamikawa**: Formal analysis (supporting); Methodology (equal). **Takeshi Yasui**: Conceptualization (equal); Funding acquisition (lead); Methodology (equal); Supervision (lead); Writing – original draft (equal); Writing – review & editing (equal).

**Data availability**

The data that support the findings of this study are available from the corresponding author upon reasonable request.

414-426 (2016).